\documentclass[twocolumn,twoside,slac_two]{revtex4}
\usepackage{graphicx}
\usepackage{fancyhdr}
\begin{document}

\title{Dark matter from extra dimensions}

%
\author{J.A.R. Cembranos}
\affiliation{Department of Physics and Astronomy, University
of California, Irvine, CA 92697, USA}
\author{A. Dobado and A.L. Maroto}
\affiliation{Departamento de F\'{\i}sica Te\'orica I,
Universidad Complutense de Madrid, 28040 Madrid, Spain}

\begin{abstract}
In brane-world models with low tension, massive branons are natural
candidates for dark matter. The phenomenology of these WIMP-like
particles is completely determined by their mass, the brane tension
and, in the case of effects due to radiative corrections, by the
cutoff setting the scale of validity of the branon effective theory.
In this paper, we review the main constraints on branon physics coming
from colliders, astrophysics and cosmological observations, and include
more recent limits obtained from electroweak precision measurements.
\end{abstract}

\maketitle
\input epsf
\thispagestyle{fancy}


\section{Introduction}
In brane-world (BW) models \cite{ADD}, the possibility that the
gravity scale is much lower than the Planck scale and
possibly close to the TeV range can give rise to interesting
observable effects at present or near
experiments \cite{Col}. The main idea that defines the
BW scenario is that the Standard Model (SM) particles are
restricted to a three-dimensional hypersurface or 3-brane,
 whereas the gravitons can propagate along the whole bulk space
 (see for example \cite{Par} for a particular construction).

Since rigid objects do not exist in relativistic theories, it
is clear that brane fluctuations must play an important role in
this framework \cite{DoMa}. This fact turns out to be
particularly true when the brane tension scale $f$
($\tau=f^4$ being the brane tension) is much smaller than the
$D$ dimensional or
fundamental gravitational scale $M_D$, i.e. $f<<M_D$. In this
case the only relevant low-energy modes of the BW scenarios are
the SM particles and branons which are the quantized brane
 oscillations. Indeed branons can be understood as the
(pseudo-)Goldstone
 bosons corresponding to the spontaneous breaking of translational
 invariance in the bulk space produced by the presence of the brane.

The branon properties allow to solve some of the problems
of the brane-world scenarios such as the
divergent virtual contributions from the Kaluza-Klein tower at the
tree level or non-unitarity of the graviton production
cross-sections \cite{GB}.
The SM-branon low-energy effective Lagrangian reads
\cite{DoMa,BSky,ACDM}:
\begin{eqnarray}
{\mathcal L}_{Br}&=&
\frac{1}{2}g^{\mu\nu}\partial_{\mu}\pi^\alpha
\partial_{\nu}\pi^\alpha-\frac{1}{2}M^2\pi^\alpha\pi^\alpha
\nonumber  \\
&+&
\frac{1}{8f^4}(4\partial_{\mu}\pi^\alpha
\partial_{\nu}\pi^\alpha-M^2\pi^\alpha\pi^\alpha g_{\mu\nu})
T^{\mu\nu}_{SM}\label{lag}
\end{eqnarray}
We see that branons interact by pairs with the SM
energy-momentum tensor. This means that they are stable particles.
On the other hand, their couplings are suppressed by the
brane tension $f^4$, i.e. they are weakly interacting. These features
make them natural dark matter \cite{CDM,M} candidates (see \cite{MaRa}
for updated reviews on cosmology and dark matter).

\section{Branon signals in colliders}
The branon signals in colliders depend on their number
$N$, the brane tension scale $f$, and their masses 
$M$. From the effective action given in the Equation (1), one can calculate
the relevant cross-sections for different branon searches.
The single photon channel and the monojet production are the more
interesting ones.
\begin{figure}[h]
\begin{center}
\resizebox{7cm}{!}{\includegraphics{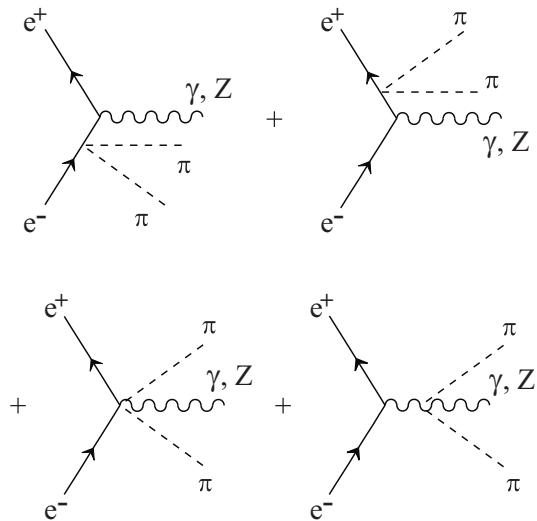}}
\caption {\footnotesize Relevant Feynman diagrams for the branon
contribution to the
single Z and the single photon channel in electron-positron colliders.} \label{LC}
\end{center}
\end{figure}
The main results in relation with this analysis are presented
in Table I, where one can find not only the present restrictions
coming from HERA, Tevatron and LEP-II but also the
prospects for future colliders like ILC, LHC or CLIC \cite{ACDM,L3,CrSt}.
\begin{table}[h]
\centering \small{
\begin{tabular}{|c|cccc|}
\hline\hline Experiment
&
$\sqrt{s}$(TeV)& ${\mathcal
L}$(pb$^{-1}$)&$f_0$(GeV)&$M_0$(GeV)\\
\hline
%
%
HERA$^{\,1}
$& 0.3 & 110 &  16 & 152
\\
Tevatron-I$^{\,1}
$& 1.8 & 78 &   157 & 822
\\
Tevatron-I$^{\,2}
$ & 1.8 & 87 &  148 & 872
\\
LEP-II$^{\,2}
$& 0.2 & 600 &  180 & 103
\\
\hline
Tevatron-II$^{\,1}
$& 2.0 & $10^3$ &  256 & 902
\\
Tevatron-II$^{\,2}
$& 2.0 & $10^3$ &   240 & 952
\\
ILC$^{\,2}
$& 0.5 & $2\times 10^5$ &  400 & 250
\\
LHC$^{\,1}
$& 14 & $10^5$ &  1075 & 6481
\\
LHC$^{\,2}
$& 14 & $10^5$ &   797 & 6781
\\
CLIC$^{\,2}
$& 5 & $10^6$ &  2640 & 2500
\\
\hline\hline
\end{tabular}
} \caption{\footnotesize{Summary of the main analysis related to 
direct branon searches in collider experiments. All
 the results are
performed at the $95\;\%$ c.l. Two different channels have been studied: the
one marked with an upper index $^1\,$ is related to monojet
production, whereas the single photon is labelled with an upper
index $^2\,$. The table contains a total of seven experiments: HERA, LEP-II, the I
and II Tevatron runs, ILC, LHC and CLIC. Obviously the data corresponding to the four
last experiments are estimations, whereas the first three analysis have been performed
 with real data.
$\sqrt{s}$ is the center of mass energy associated to the total
process; ${\mathcal L}$ is the total integrated luminosity;
 $f_0$, the bound in the brane tension scale for one
massless branon ($N=1$) and $M_0$ the limit on the branon mass for 
small tension $f\rightarrow0$.}}
\label{tabHad}
\end{table}

\section{One loop effects}

In addition to direct production and the corresponding missing
energy signatures, branons can also give rise to new effects
through radiative corrections. By integrating out the branon
fields in the action coming from ${\mathcal L}_{Br}$ it is
possible to obtain an effective action for the SM particles which
includes the effect produced by branon loops. At the level of
two-point functions, branon loops result only in a renormalization
of the SM particle masses which is not observable. However new
couplings appear which can be described by an effective
lagrangian \cite{radcorr} whose more relevant terms are:
\begin{eqnarray}
{\mathcal L}_{eff}= W_1 T_{\mu\nu}T^{\mu\nu}+W_2 T_\mu^\mu
T_\nu^\nu  \label{eff}.
\end{eqnarray}
where  $T^{\mu\nu}\equiv T^{\mu\nu}_{SM}$ and
\begin{eqnarray}
W_1 &=& \frac{N \Lambda^4}{96(4\pi)^2f^8}
\nonumber  \\
W_2 &=& \frac{N \Lambda^4}{192(4\pi)^2f^8}
\end{eqnarray}
for $\Lambda>>M$, $\Lambda$ being  the cutoff setting the limit of
validity on the effective description of branon and SM dynamics
used here. This new parameter appears when dealing with branon
radiative corrections since the lagrangian in (\ref{lag}) is not
renormalizable. When the branon mass $M$ is not small compared
with $\Lambda$, $ W_1$ and  $ W_2$ have much more involved
definitions, which will be given elsewhere \cite{radcorr}.

\begin{center}
\begin{figure}[h]
\begin{tabular}{c c c}
\begin{tabular}{c c c}
$\psi_a(p_2)$& &$\psi_b(p_4)$\\
 &\epsfysize=1 cm\epsfbox{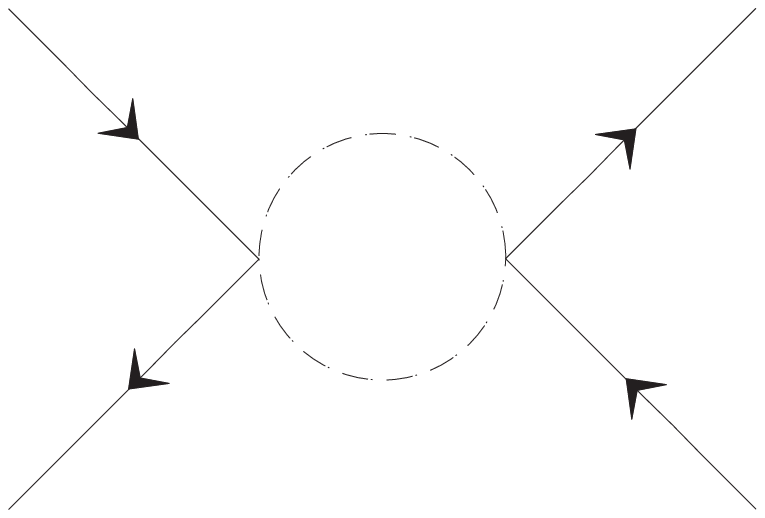}& \\
 $\bar\psi_a(p_1)$& &$\bar\psi_b(p_3)$
\end{tabular}
 &
$\Rightarrow$
&
\begin{tabular}{c c c}
$\psi_a(p_2)$& &$\psi_b(p_4)$\\
 &\epsfysize=1 cm\epsfbox{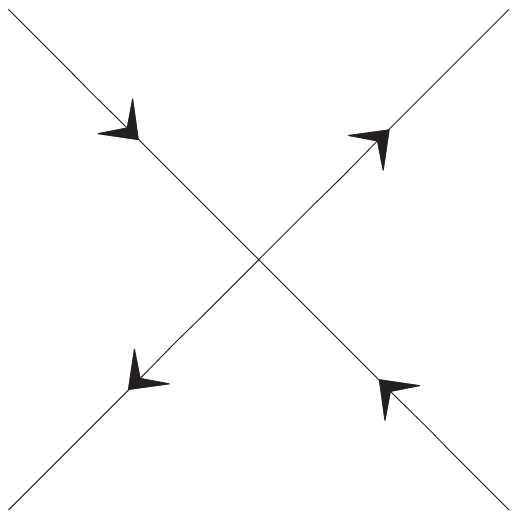}& \\
 $\bar\psi_a(p_1)$& &$\bar\psi_b(p_3)$
\end{tabular}
\end{tabular}
\caption {\footnotesize Four-fermion vertex induced by branon radiative corrections.}
\label{loop}
\end{figure}
\end{center}

An effective lagrangian similar to the one in (\ref{eff}) was
obtained in \cite{CrSt,GS} by integrating at the tree level the
Kaluza-Klein modes of gravitons propagating in the bulk and some
of its phenomenological consequences where studied there. Thus it
is easy to translate some of the results from these references to
the present context. For example one of the most relevant
contributions of  branon loops to the SM particle phenomenology
could be the four-fermion interactions appearing in (\ref{eff})
(see Fig. \ref{loop}) or fermion pair annihilation into two gauge bosons.

Following  \cite{GiSt} it is possible to use the data
coming from LEP, HERA and Tevatron on this kind of processes  to
set bounds on the parameter combination $f^2/(\Lambda N^{1/4})$.
The results are shown in Table \ref{constrains1}. It is
interesting to see that the various constraints found are not
too different.

\

\begin{table}
\centering
\begin{tabular}{|c|c|c|}
\hline\hline Experiment&Process & $f^2/(\Lambda
N^{1/4})$ (GeV)\\
\hline
LEP combined~\cite{lepc} & $\gamma \gamma$ & 59\\
LEP combined~\cite{lepcc} & $e^+e^-$ & 75\\
H1~\cite{h1} & $e^+p$ and $e^-p$ & 47\\
ZEUS~\cite{zeus} & $e^+p$ and $e^-p$ & 46\\
D\O~\cite{d0c} & $e^+e^-$ and $\gamma \gamma$ & 69\\
CDF~\cite{zhou} &  $e^+e^-$ and $\gamma \gamma$ & 55\\
\hline
\multicolumn{2}{|c|}{combined~\cite{GiSt}}&81\\
\hline\hline
\end{tabular}
\caption{\label{constrains1} \footnotesize{Estimated lower limits on
$f^2/(N^{1/4}\Lambda)$ (in GeV) provided from different
experiments.}}
\end{table}

In a similar way, using the analysis in \cite{Hewett}, it is
possible to estimate the constraints that could be found in the
next generation of colliders. For that purpose,  we have taken 
into account the
estimations calculated by Hewett  for future linear colliders like
the ILC, the  Tevatron run II and the LHC (see Table
\ref{Future}).

\begin{table}
\centering
\begin{tabular}{|c|c|c|c|} \hline\hline
  & $\sqrt s$ (TeV) & ${\cal L}$ (fb$^{-1}$) & $f^2/(N^{1/4}\Lambda)$ (GeV) \\ \hline
ILC     & 0.5   & 75  & 216 \\
       & 0.5   & 500 & 261  \\
       & 1.0   & 200 & 421  \\ \hline
Tevatron II & 1.8 & 0.11 & 63 \\
         & 2.0 & 2    & 83 \\
         & 2.0 & 30   & 108 \\ \hline
LHC      & 14  & 10   & 332 \\
         & 14  & 100  & 383 \\ \hline\hline
\end{tabular}
\caption{\footnotesize{Estimated constraints on the parameter
$f^2/(\Lambda N^{1/4})$ in GeV for some future colliders.
$\sqrt{s}$ is the center of mass energy associated to the total
process and ${\mathcal L}$ is the total integrated luminosity.}}
\label{Future}
\end{table}

\section{Electroweak precision observables and anomalous magnetic moment}

Electroweak precision measurements are very useful to constrain
models of new physics. The so called oblique corrections
(the ones corresponding to the $W$, $Z$ and $\gamma$ two-point
functions) use to be described in terms of the $S,T,U$ \cite{STU}
or the $\epsilon_1,\epsilon_2$ and $\epsilon_3$ parameters
\cite{Alta}. The first order correction coming from the
Kaluza-Klein gravitons in the ADD models for rigid branes to the
parameter:
\begin{equation}
 \bar\epsilon\equiv
\frac{\delta M_W^2}{M_W^2} - \frac{\delta M_Z^2}{M_Z^2}
\end{equation}
was computed in \cite{CPRS}. Translating this result to our
context as in the previous section we find:
\begin{equation}
  \delta \bar\epsilon \simeq
   \frac{5\,(M_Z^2-M_W^2)}{12\,(4\pi)^4}
  \frac{ N\Lambda^6}{f^8}
\end{equation}

The experimental value of $\bar\epsilon$ obtained from LEP \cite{EWWG}
is $\bar\epsilon=(1.27 \pm 0.16)\times 10^{-2}$. This value is consistent with
the SM prediction for a light higgs $m_H\leq 237$ GeV at 95 \% c.l.
On the other hand, the theoretical uncertainties are one order of magnitud smaller
\cite{Alta} and therefore, we can estimate the constraints for the branon contribution
at 95 \% c.l. as $|\delta\bar\epsilon|\leq 3.2\times 10^{-3}$.
Thus it is possible to set the bound:
\begin{equation}\label{eb}
    \frac{f^4}{ N^{1/2}\Lambda^3}\geq\;3.1\; 
\mbox{GeV}\; (\,95\; \%\; c.l.\,)
\end{equation}
This result has a
stronger dependence on $\Lambda$ ($\Lambda^6$) than the
interference cross section between the branon and SM interactions
($\Lambda^4$). Therefore, the constraints coming from this analysis are
complementary to the previous ones.

A further constraint to the branon parameters can be obtained  
from the  $\mu$ anomalous magnetic moment.  The first branon contribution
to this parameter can be obtained  from a one loop computation with the lagrangian
 given by (\ref{eff}).

The result for the KK graviton tower was first calculated  by
\cite{Graesser} and confirmed by \cite{CPRS} in a different way
and can be written as:

\begin{equation}
\delta a_\mu \approx \frac{2 m_\mu^2 \Lambda^2}{3(4\pi)^2}
  (11\,W_1-12\,W_2),
\end{equation}

which for the branon case can be written as:
\begin{equation}\label{gb}
\delta a_\mu \approx \frac{5\, m_\mu^2}{114\,(4\pi)^4}
  \frac{N\Lambda^6}{f^8}.
\end{equation}
This result depends on the cut-off $\Lambda$ in the same way as 
the electroweak precision parameters. However the experimental
situation is a little different. In a sequence of increasingly
more precise measurements, the  821 Collaboration at the
Brookhaven Alternating Gradient Syncrotron has reached a fabulous
 relative precision of 0.5 parts per million in the determination
 of $a_\mu=(g_\mu-2)/2$ \cite{BNL}. These measurements provide a
 stringent test not only of new physics but also to the SM. Indeed,
 the present result is only marginally consistent with the SM. Taking
 into account the $e^+e^-$ collisions to calculate the $\pi^+\pi^-$
 spectral functions, the deviation with respect to
the SM prediction is at
 $2.6$ standard deviations \cite{gm2}. In particular: $\delta a_\mu \equiv a_\mu
(exp) - a_\mu (SM) =(23.4 \pm 9.1)\times 10^{-10}$.
Using Equation (\ref{gb}) we can estimate the {\it preferred} parameter region for branons
 to provide the observed difference:
\begin{equation}
    6.0\; \mbox{GeV}\;\geq\,\frac{f^4}{ N^{1/2}\Lambda^3}\,
\geq\;2.2\; \mbox{GeV}\; (\,95\; \%\; c.l.\,)
\end{equation}
We observe that the correction to the muon anomalous
magnetic moment is in the right direction and that
 it is possible to avoid the present constraints and
 improve the observed experimental value by the E821 Collaboration.
 
There are two interesting comments related to these results. 
First if there
 is new physics in the muon anomalous
magnetic moment and this new physics is due to branon radiative
 corrections, the phenomenology
of these particles should be observed at the LHC
and in a possible future ILC (see Table \ref{Future}). In particular,
the LHC should observe an important difference in the channels: $pp\rightarrow e^+e^-$ and
 $pp\rightarrow \gamma \gamma$
with respect to the SM prediction. The ILC should observe the most important effect in
 the process:
$e^+e^-\rightarrow e^+e^-$.

On the other hand, it is interesting to note that the same physics 
that could explain the
Dark Matter content of the Universe could also explain the magnetic 
moment deficit of the muon.
In fact, as we show below, the above branon models with order of 
magnitude masses between $M\sim 100$ GeV and $M\sim 10$ TeV present the total
non baryonic Dark Matter abundance observed by different 
experiments \cite{CDM,MaRa}.
In such a case, the first branon signals at colliders would 
be associated to the radiative
corrections described in this section \cite{radcorr} and not to the direct 
production studied in previous works
\cite{ACDM}.

\section{Cosmological and astrophysical limits}

\begin{figure}[h]
\begin{center}
\resizebox{8cm}{!}{\includegraphics{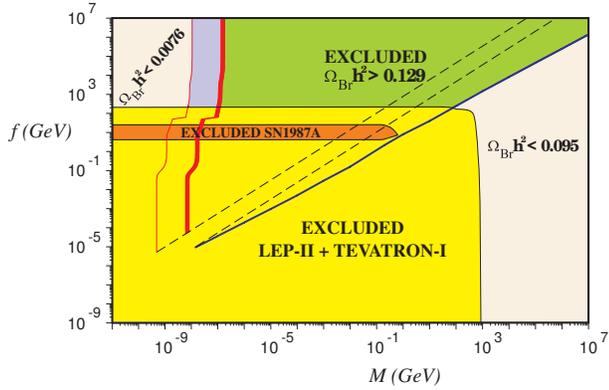}}
\caption {\footnotesize Relic abundance in the $f-M$ plane for a model with one branon of
mass $M$. The two lines on the left correspond to the
$\Omega_{Br}h^2=0.0076$ and $\Omega_{Br}h^2=0.129 - 0.095$ curves
for hot-warm relics, whereas the right line corresponds to the
latter limits for cold relics (see \cite{CDM} for details). The
lower area is excluded by single-photon processes at LEP-II
\cite{ACDM,L3} together with monojet signal at Tevatron-I
\cite{ACDM}. The astrophysical constraints are less restrictive
and they mainly come from supernova cooling by branon emission
\cite{CDM}.} \label{mother}
\end{center}
\end{figure}

The potential WIMP nature of branons means that
these new particles are natural dark matter candidates.  In
\cite{CDM} the relic branon abundance has been calculated
in two cases: either
relativistic branons at freeze-out (hot-warm) or non-relativistic
(cold), and assuming that the evolution of the universe is
standard for $T<f$ (see Fig. \ref{mother}).
Furthermore, if the maximum
temperature reached in the universe is smaller than the branon
freeze-out temperature, but larger than the explicit symmetry
breaking scale, then branons can be considered as massless
particles decoupled from the rest of matter and radiation. In such
a case, branons can act as nonthermal relics and, in the particular
case in which the total number of dimensions is six, it is possible to
relate the cosmic
coincidence problem with the existence of large extra dimensions
\cite{M}.

%
\begin{figure}[ht]
\begin{center}
\resizebox{8cm}{!}{\includegraphics{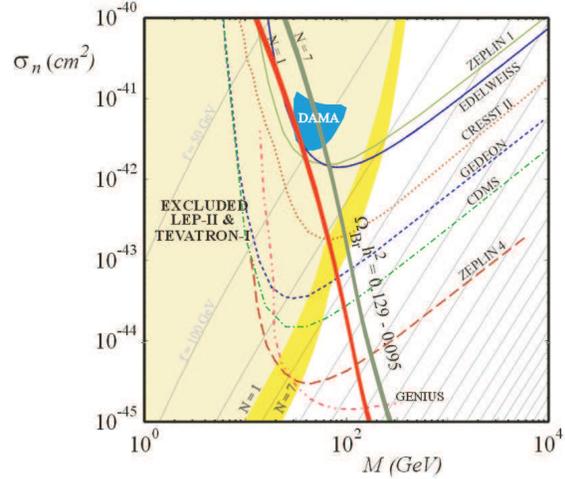}}
\caption {\footnotesize Elastic branon-nucleon cross section $\sigma_n$
 in terms
of the branon mass. The thick (red) line corresponds to the
$\Omega_{Br}h^2=0.129-0.095$ curve for cold branons in Fig. 2
from $N=1$ to $N=7$. The shaded areas are the LEP-II
and Tevatron-I exclusion regions. The
solid lines correspond to the current limits on the
spin-independent cross section from direct detection experiments.
The discontinuous lines are the projected limits for future
experiments. Limits obtained from \cite{dmtools}.} \label{direct}
\end{center}
\end{figure}

If branons make up the galactic halo, they could be detected by
direct search experiments from the energy transfer in elastic
collisions with nuclei of a suitable target. From
Fig. \ref{direct} we see that if branons constitute the
dominant dark matter component, they could not be
detected by present experiments such as DAMA, ZEPLIN 1 or
EDELWEISS. However, they could be observed by future detectors
such as CRESST II, CDMS or GENIUS \cite{CDM}.

Branons could also be detected indirectly: their annihilations in
the galactic halo can give rise to pairs of photons or $e^+ e^-$
which could be detected by $\gamma$-ray telescopes such as MAGIC
or GLAST or antimatter detectors (see \cite{CDM} for an estimation
of positron and photon fluxes from branon annihilation in AMS).
Annihilation of branons trapped in the center of the sun or the
earth can give rise to high-energy neutrinos which could be
detectable by high-energy neutrino telescopes such as AMANDA,
IceCube or ANTARES. These searches complement those already
commented in high-energy
particle colliders (both in $e^+ e^-$ and hadron colliders
\cite{L3,ACDM}) in which real (see Fig. \ref{mother}) and virtual branon
effects could be measured. Finally, quantum fluctuations
of branon fields during inflation can give rise to CMB anisotropies
through their direct contribution to the induced metric (work is
in progress in these directions).

\section{Conclusions}

In this paper we have reviewed the main features of branon physics.
We have considered their main phenomenological signals parametrized
in terms of the branon mass, the brane tension scale and the
effective theory
cutoff scale. At the tree level, the
most important signals come from missing energy and momentum events
in single photon, single Z and monojet processes in colliders. 
At one-loop level, the most interesting processes are those
involving new four fermion interactions and finally, at the
two-loop level,  the electroweak
precision measurements and the muon anomalous magnetic moment 
can also set bounds on the model parameters. We have also considered
the limits coming from cosmology  due to the WIMP-like
nature of branons. This, in turn, ensures the existence of a relic 
abundance of branons, which could make up the galactic haloes.   

\section*{Acknowledgments}
This work is partially supported by DGICYT (Spain) under project numbers
FPA 2000-0956 and BFM 2002-01003 and by the Fulbright-MEC (Spain) program.
A.D. acknowledges  the hospitality of the SLAC Theory Group,
where the final part of this work was done and economical support
from the Universidad Complutense del Amo Program.


\end{document}